\documentstyle[psfig]{caosp}
\def\infig#1#2#3{\epsfxsize=#3cm \centering{\mbox{\epsfbox{#2}}}\vspace{-0.4cm}}
\def\two{\hbox{{$\scriptstyle {\rm II}$}}}
\def\kms{km\,s$^{\mathrm -1}$~}

\begin{document}

\title{Linear spectropolarimetry of Ap stars: a new degree of constraint on
magnetic structure}

\author { G.A. Wade\inst{1}, J.-F. Donati\inst{2}, G. Mathys\inst{3},
N. Piskunov\inst{4}}
\institute{Physics \& Astronomy Department, University of Western Ontario,
London, Ontario, Canada, N6A 3K7
\and
Observatoire Midi-Pyr\'en\'ees, 14 Avenue Edouard Belin, 31400 Toulouse,
France\and ESO, Casilla 19001, Santiago 19, Chile
\and
Uppsala Astronomical Observatory, Box 515, S-751 20 Uppsala, Sweden}
\maketitle

\begin{abstract}
We present preliminary results from a programme aimed at acquiring
 linear spectropolarimetry of magnetic A and B stars. Linear polarization
 in the spectral lines of these objects is due to the Zeeman effect,
 and should provide detailed new information regarding the structure
 of their  strong magnetic fields.  To illustrate the impact of these
 new data, we compare observed circular and linear polarization
 line profiles of 53~Cam with the profiles predicted by the magnetic
 model by Landstreet (1988). 

 Linear polarization in the spectral lines of {\em all} stars studied
 is extremely weak; in most cases, below the threshold of
 detectability even for very high SNRs. In order to overcome this
 problem, we employ the Least-Squares Deconvolution (LSD) multi-line
 analysis technique in order to extract low-noise mean line profiles and
 polarization signatures from our \'echelle spectra. Tests show that
 these mean signatures can be modelled as real spectral lines, and
 have the potential to lead to high-resolution maps of the magnetic
 and chemical abundance surface distributions.

\end{abstract}


\section{Introduction}

Recent observations of the broadband linear polarization variability
of magnetic Ap stars
have been shown to be very sensitive to the
detailed topology of their magnetic fields. Using these data, Leroy et
al. (1996) have presented models of the magnetic field structure of
several stars which resolve clear departures from the simple
multipolar geometries assumed in the past. Following these studies, we
have taken an important next step by acquiring systematic linear {\em
spectropolarization} measurements of a number of strong-field Ap
stars. Due principally to the additional information contained in line
profile shapes, these data promise important new constraints on the
structure of the magnetic fields of these objects, and also offer the
possibility of constructing mutually consistent, detailed magnetic
field and chemical abundance distribution maps for individual stars. This will
in turn provide new information about the interaction of the magnetic
field with the atmospheric processes responsible for abundance
inhomogeneities, as well as motivate new theoretical investigation
into the mechanisms responsible for magnetic fields in early-type
stars.

\section{Spectropolarimetric observations}

 Using the new MuSiCoS spectropolarimeter and ESO CASPEC
 spectropolarimeter,  we have acquired multiple observations of a
 number of strong-field magnetic A and B stars in each of the four
 Stokes parameters. Fig. 1 shows that, in the case of the cool Ap star
 $\beta$~CrB, circular polarization (Stokes V) and linear polarization
 (Stokes Q) are clearly detected in individual spectral
 lines. $\beta$~CrB is unusual in this respect; most stars studied
 display linear polarization amplitudes which are too weak to be
 detected in our high SNR ($\geq 300$)  spectra.  However, it turns
 out that even marginal detections or upper limits of  the linear
 polarization can provide some surprising new results.

\subsection{First results for 53~Cam}

 The magnetic field and chemical abundance distributions of the Ap star
 53~Cam were modelled
 by Landstreet (1988). This model makes specific predictions about the
 shapes and amplitudes of the circular and linear polarization
 profiles of individual spectral lines at any rotational phase. In
 Fig. 2 we compare the observed Stokes I, V, Q and U profiles of
 Fe~\two~$\lambda 4923.9$ on 1997 February 16 with those predicted for
 this line at this phase by  Landstreet's model. The agreement is not
 very good; the amplitude of the circular polarization (Stokes V)
 profile is overestimated by a factor of 2, while the linear
 polarizations (Stokes Q and U) are respectively  overestimated by
 factors of 2.8 and at least 7.7. 
 Similar behaviour is observed throughout the entire 
 rotation of the star, at six other observed rotational phases. This
 represents a serious disagreement, and indicates that these new data
 will provide important new refinements to the best magnetic field
 models currently available.  

\vspace{-2mm}
\section{Least-Squares Deconvolution}

 Our inability to detect linear polarization in the spectral lines of most
 stars studied indicates that 
 the linear polarization profiles must be extremely weak. While marginal
 detections and upper limits are sufficient for illustrations such as that
 presented above, detailed modeling requires a reasonable relative SNR for the
 polarization profiles.
 To usefully detect the weak linear polarizations and to obtain a higher 
 relative SNR, we
 exploit the similar information contained in the many spectral lines
 found in our \'echelle spectra using the Least-Squares Deconvolution
 (LSD) technique (Donati et al. 1997).

\vspace{-2mm}
\subsection{Technique}

 LSD is a technique for the simultaneous analysis of many spectral
 lines. LSD assumes that all
 spectral features in a given spectrum (Stokes I, V, Q or U) are
 identical in shape - they differ only in amplitude by known scaling
 factors. This assumption implies that the stellar spectrum is effectively the
 convolution of a shape function (called the  {\em mean signature})
 with a ``spectrum'' of weighted delta functions (called the {\em line
 mask}). LSD consists of deconvolving the 
 mean signature from the spectrum, given knowledge of the line mask.
 
 Of course, the  principal assumption is not strictly true
 (the scaling relations employed are strictly valid only for weak lines
 and weak magnetic fields; shapes of different spectral lines are not 
 identical: Zeeman splitting
 patterns vary from line to line, thermal width of 
 lines varies with atomic weight, etc.), and it is not clear {\em a
 priori} that LSD should produce useful results. We have therefore
 performed realistic tests with synthetic spectra to  determine to
 what degree the mean signatures produced by LSD resemble
 representative individual spectral lines.

\vspace{-2mm}
\subsection{Tests}

 A synthetic spectrum was computed
 using VALD linelists for a known surface distribution of
 magnetic field and chemical abundances. The spectrum was
 convolved with a gaussian instrumental profile with a FWHM of
 8~\kms and infected with random noise in order to generate a SNR of 250:1.
 In this way the spectra were made to simulate the real data obtained
 by the MuSiCoS spectropolarimeter. For all but the strongest lines in
 this spectrum the linear polarization signatures were below the level
 of the noise, similar to the situation encountered in the observed spectra.

 Finally, the spectrum was Least-Squares Deconvolved and mean signatures
 (in each of the Stokes I, Q, U and V parameters) were extracted. 
 How well do these mean signatures agree with a ``representative''
 spectral line (assumed, as a first approximation, to be a pure Zeeman triplet
 with a depth and Land\'e factor equal to the mean of all lines used) computed
 assuming identical surface distributions of chemical abundance and magnetic
 field? In Fig. 3 the mean signatures extracted from the
 synthetic spectrum are compared with the ``representative'' spectral
 line described above. The agreement is
 excellent, and indicates that the mean signatures extracted by LSD can
 be modelled as real spectral lines, assuming a line profile model
 for which the parameters are known {\em a priori}. As well, the relative
 SNR of the LSD mean signatures is very high, allowing for detailed modeling
 which would not have been possible using individual lines in the original
 spectrum.  
\twocolumn
\def\infig#1#2#3{\epsfxsize=#3cm \centering{\mbox{\epsfbox{#2}}}\vspace{-0.4cm}}
\def\two{\hbox{{$\scriptstyle {\rm II}$}}}
\renewcommand{\topfraction}{1.0}
\renewcommand{\bottomfraction}{1.0}
\renewcommand{\textfraction}{0.0}
\renewcommand{\floatpagefraction}{1.0}
\setcounter{topnumber}{5}
\setcounter{bottomnumber}{5}
\setcounter{totalnumber}{10}
\begin{figure}[hbtp] 
\psfig{figure=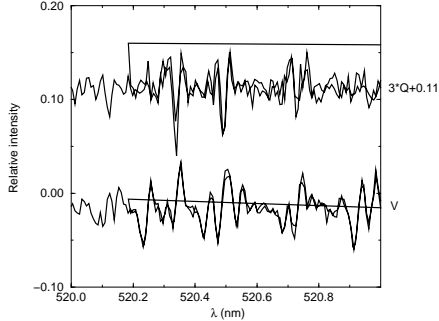,height=5cm} \caption{Stokes V
(bottom) and Q (top) spectra of $\beta$~CrB, obtained using the MuSiCoS
spectropolarimeter. Each spectrum shows two overlapping \'echelle orders.
Both V and Q are clealy detected above the noise level.} 
\end{figure}
\begin{figure}[hbtp] 
\psfig{figure=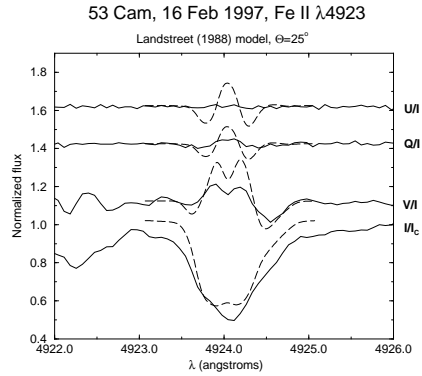,height=5cm}
\caption{{\em Solid curves (bottom to top)} -- Stokes I, V, Q and U
profiles of Fe~\two~$\lambda 4923.9$ for the Ap star 53~Cam, observed on
16 February 1997. {\em Broken curves} -- Predictions of Landstreet's
(1988) model. The amplitudes of all polarizations are severely
overestimated by the model (by a factor of 2 for Stokes V, by a factor of
4.3 for Stokes Q, and by at least a factor of 8.3 for Stokes U).}
\end{figure}
\begin{figure}[htp]
\psfig{figure=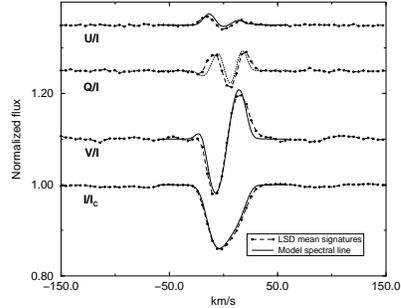,height=4.5cm}
\caption{Tests of LSD using synthetic spectra: LSD mean signatures are
extracted from a synthetic spectrum, and then compared for fidelity with
the spectrum. Here, LSD mean signatures are compared with a
synthetic spectral line with identical mean parameters.
% Even in this
% preliminary example, it is clear that the LSD signatures agree quite well
% with the original. In particular, it should be noted that
 The amplitudes
and shapes of the circular and linear polarization profiles are extremely
well reproduced.} 
\end{figure}
\begin{figure}[hbtp]
\psfig{figure=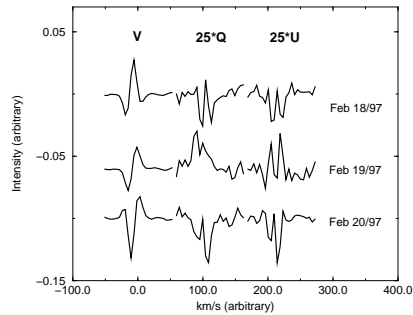,height=4.5cm}
\caption{LSD Stokes V, Q and U mean signatures for the Ap star 78~Vir,
observed on successive nights in Feb. 1997.
For clarity the linear polarization profiles have been magnified
by $25\times$, although they are clearly detected above the noise.} 
\end{figure}
\onecolumn
\subsection{Results}
 LSD has been employed to extract mean Stokes signatures from polarization
 spectra of more than 10 magnetic Ap stars. Since most of these stars do
 not display a sufficiently large linear polarization amplitude to allow its
 detection in our spectra, LSD provides a unique
 opportunity to characterize and model the line profile linear polarization.
 In Fig. 4 we show LSD mean Stokes signatures for the Ap star 78~Vir, 
 obtained over three nights in February 1997. Clear rotational modulation
 of all three Stokes parameters is visible. By obtaining polarization 
 spectra of individual stars with good sampling of the entire rotational
 cycle, mutually consistent magnetic field and chemical abundance maps can be 
 constructed using a mapping code such as INVERS10 (Piskunov, 
 these proceedings). Such maps will comprise important contributions to our
knowledge of the structure of the magnetic fields of A and B stars, as well
as to our understanding of how the transport processes responsible for
chemical peculiarities are affected by the presence of a magnetic field.

\section{Conclusion}

 The results presented herein show that spectral line linear polarization 
 measurements provide important new constraints on the magnetic field structure 
 of A and B stars. Least-Squares Deconvolution is shown to be an effective 
 technique for the extraction of mean Stokes signatures which can be accurately 
 modelled as real line profiles. Such modelling has the potential to lead to
the construction of mutually consistent, high-resolution maps of the magnetic 
 and chemical surface structure of these stars. 

\acknowledgements {Thanks to the Time Allocation Committees of 
l'Observatoire du Pic du Midi and the European
Southern Observatory, La Silla, for continued support of this programme.}

\end{document}